\documentclass[fleqn,twoside]{article}
\usepackage{espcrc2}

\newcommand{\AmS}{{\protect\the\textfont2
  A\kern-.1667em\lower.5ex\hbox{M}\kern-.125emS}}
\def \ii{\mathrm{i}}

\def \pd{\partial}
\def \e{\mbox{e}}
\def \lcx{\tilde{x}}

\title{Nonlocal operators and distribution amplitudes of 
definite twist, WW-- relations, BC--sum rules and
power corrections for hard QCD processes\thanks{
Contribution to Int. LC Workshop ``Light-cone Physics: Particles
and Strings'', Trento, Sept. 3 -- 11, 2001}}

\author{Bodo Geyer$^{{\rm a}}$,
	Markus Lazar
	\address{Center for Theoretical Studies and
	Institute of Theoretical Physics, Leipzig University,\\
        Augustusplatz 10, D-04109 Leipzig, Germany} %
        and
        Dieter Robaschik
	\address{Faculty 1, Brandenburg Technical University, 
	Universit\"atsplatz 3 - 4, D-03044 Cottbus, Germany}}

\begin{document}

\maketitle

\vspace{-.5cm}

Scattering amplitudes of light-cone dominated hadronic processes 
are usually represented
by the convolution of a hard (process-dependent) partonic scattering 
amplitude with appropriate soft (process-independent) distribution
amplitudes (DA). Different phenomenological DA's, as
parton distributions (PD) in deep inelastic scattering and 
in Drell-Yan processes, hadronic wave functions as well as generalized 
parton distributions (GPD), like double distributions (DD), are given as 
(non-forward) matrix elements of some nonlocal QCD light-cone operators.
Thereby, one and the same operator may be related to different hadronic
processes. Moreover, these operators by its twist decomposition also 
contribute to different DA's. Therefore, it is necessary 
to disentangle these various twist contributions. 

Conventionally, 
the notion of 
{\it dynamical} twist ($t$) is taken \cite{JJ92} by counting powers of 
the momentum transfer, $Q^{2-t}$, thereby using the (quark) equations of 
motion in the light-cone formulation. 
Being defined only for {\em matrix elements} of operators it is
easily determined experimentally by simple counting rules.
However, this phenomenological notion is {\em not} Lorentz 
invariant and, for general non-forward matrix elements, it is
{\em not uniquely defined} resulting in some arbitrariness in the
definition of GPD's. 
Furthermore, it
{\em cannot distinguish} uniquely between radiative and kinematical 
resp.~mass corrections of scattering amplitudes and, beyond
leading order, the higher dynamical twists $t$ are {\em mismatched} 
with geometric twists $\tau_i\leq t$ to be considered now.

Recently, we used a group theoretic procedure~\cite{GLR99} to decompose 
nonlocal QCD operators into a series of operators of definite {\it geometric} 
twist $\tau~=~$dimension$~d$ -- spin$~j$. 
 For the `centred' tensor 
operators, ${\cal O}_{\Gamma}(\kappa x,-\kappa x)=
\bar{\psi}(\kappa x)\Gamma U(\kappa x,- \kappa x)\psi(-\kappa x)$, 
the consecutive steps of that twist decomposition are:\\
(1) {\it Taylor expansion} of the nonlocal tensor operator 
into an infinite tower of {\em local} tensor operators 
${\cal O}_{\Gamma \mu_1 \ldots \mu_n}$ having 
rank $[\Gamma]+n$ and dimension $d$;\\
(2) {\it Decomposition} of these local operators into
{\it irreducible} ones -- with respect to the Lorentz group -- whose 
symmetry class is characterized by Young patterns
$[m]= (m_1,m_2, \ldots,m_r)$ of $r$ rows;\\
(3) {\it Resummation} of the irreducible local tensor operators belonging 
to the same symmetry class $[m]$ and having the same twist $\tau$
to a nonlocal tensor operator with definite twist.

This gives the {\em infinite off-cone} decomposition:
\begin{eqnarray}
\hspace{-.7cm}
&&{\cal O}_{\Gamma}(\kappa x,-\kappa x)
= \sum_{\tau \geq \tau_{\rm min}}
c^{(\tau)\Gamma'}_\Gamma\!\!(x)
{\cal O}^{(\tau)}_{\Gamma'}(\kappa x,-\kappa x),
\nonumber
\\
\hspace{-.7cm}
&&{\cal O}^{(\tau)}_{\Gamma}(\kappa x,-\kappa x)
= {\cal P}^{(\tau)\Gamma'}_{~\Gamma}\!\!(x,\pd)
{\cal O}_{\Gamma'}(\kappa x,-\kappa x),
\nonumber
\\
\hspace{-.7cm}
&&\big({\cal P}^{(\tau)} \times
{\cal P}^{(\tau')}\big)^{~\Gamma'}_{\Gamma}
= \delta^{\tau \tau'}{\cal P}^{(\tau)\Gamma'}_{~\Gamma},
\nonumber
\end{eqnarray}
where ${\cal P}^{(\tau)}_{\Gamma\Gamma'}$ are  well-defined projection
operators, and
$c^{(\tau)\Gamma'}_\Gamma\!\!(x)$ are coefficient functions 
depending on (powers of) $x^2$ and, according to the
tensor structure $\Gamma$, also on $x_\mu$ and/or $\pd_\mu$.\\
(4) {\it Projection} onto the light--cone,
$x \rightarrow \lcx$ with $\lcx^2=0$, results in a {\em finite on-cone} 
decomposition because then only a finite number of the coeffitient functions
survive.

The scalar operator 
${\bar\psi}(\kappa x) x\gamma \psi(-\kappa x)$, for example,  
has the following infinite decomposition:
\begin{eqnarray}
\hspace{-.7cm}
&&
O(\kappa x,-\kappa x)=O^{{\rm tw}\,2}(\kappa x,-\kappa x)
+\sum_{j=1}^{\infty}
\frac{(-1)^j (\kappa x)^{2j}}{4^j j!(j-1)!}
\nonumber\\
\hspace{-.7cm}
&& \qquad\quad \times
\int_0^1 d t\, t\, (1-t)^{j-1}\, 
O^{{\rm tw}(2+2j)}(\kappa tx,-\kappa tx),
\nonumber
\end{eqnarray}
with the harmonic operators of twist 
$\tau=2+2j$:
\begin{eqnarray}
\hspace{-.7cm}
&&
O^{{\rm tw}(2+2j)}(x,-x)
=\int\!d q\, 
\left(\bar\psi \gamma^\mu \psi \right)\!(q)\,(- q^2)^{j}\Big\{\!
\big[ x_\mu
\nonumber\\
\hspace{-.7cm}
&&
\times(2+q\pd_q)
-\hbox{\large$\frac{1}{2}$}\, {\ii} q_\mu x^2  
\big]
(3+q\pd_q)
{\cal H}_2\big((qx)^2\!-\!q^2\! x^2\big)
\nonumber\\
\hspace{-.7cm}
&&\qquad
\; - 2j\, \frac{ {\ii} q_\mu}{ q^{2}} \left(1+q\pd_q\right)
{\cal H}_1\big((qx)^2\!-\!q^2\! x^2\big)
\Big\}\,\e^{{\ii} qx/2},
\nonumber
\end{eqnarray}
where $
{\cal H}_\nu(z)=
\sqrt{\pi}\left(\sqrt{z}\right)^{1/2-\nu}
\!J_{\nu-1/2}\left(\hbox{\large$\frac{1}{2}$}\sqrt{z}\right)
$.

In contrast to dynamical twist the geometric twist is a 
Lorentz invariant {\em notion for operators} and, being a `good' quantum
number, is conserved under renormalization. The 
matrix elements of ${\cal O}^{(\tau)}_{\Gamma}(\kappa x,-\kappa x)$
define general (multi-variable) DA's
of definite twist $\tau$:\\
$\bullet$~~forward matrix elements on the light-cone are related to new 
{\em twist-$\tau$ parton distributions}~\cite{GL01}, \\
$\bullet$~non-forward off-cone matrix elements lead to new 
{\em generalized twist-$\tau$ distribution amplitudes}  
and their {\em power resp.~mass corrections}~\cite{GLR01},\\
$\bullet$~new {\em twist-$\tau$ meson distribution amplitudes} are obtained
as special cases~\cite{L01,GLR01}.

The price for the field theoretical advantages of the
new GPD's 
is that they are experimentally 
not as easily accessible as the conventional ones are. 
However, both kinds of GPD's are {\em uniquely related} and,
from these relations, without use of equations of motion, 
one obtains~\cite{GL01}\\
$\bullet$~Wandzura-Wilczek--like integral relations, and \\
$\bullet$~Burkhardt-Cottingham--like sum rules,\\
for the conventional (quark) distributions of dynamical twist;
this generalizes also to  multi-variable GPD's 
and  meson DA's, for the latter, see,~\cite{L01}.

Before introducing double distributions of definite twist let us consider 
the most general parametrization of non-forward matrix elements:
\begin{eqnarray}
\hspace{-.7cm}
&&
\langle P_2,S_2 |{\cal O}_{\Gamma}(\kappa x,-\kappa x)|P_1,S_1 \rangle
\nonumber\\
\hspace{-.7cm}
&&= 
{\cal K}_\Gamma^a({\cal P})
\int {D}{\cal Z}\, 
\e^{{\ii}\kappa (x{\cal P}){\cal Z}}\,
f_a({\cal Z}, {\cal P}_i {\cal P}_j, x^2; \mu^2 ),
\nonumber
\end{eqnarray}
where the following notations have been used:\\
${\cal P} = \{P_+, P_-\}$, ${\cal Z} = 
\{z_+,  z_-\}$ with  $P_\pm = P_2\pm P_1$, 
$z_\pm=\hbox{\large$\frac{1}{2}$} (z_2 \pm z_1)$,
${D}{\cal Z}= dz_1 dz_2 \theta(1-z_1)\theta(z_1+1)
\theta(1-z_2)\theta(z_2+1)$.
The support restriction, 
$-1\leq z_i \leq 1$,  occurs since
the matrix elements are {\em entire analytic functions} r.w.t.~$(x P_i)$.\\
 ${\cal K}_\Gamma^a({\cal P})$ are {\em linear independent spin
structures} being defined by the help of (free) hadron wave functions, 
e.g., for virtual Compton scattering the Dirac and Pauli structures, 
$
\bar u(P_2,S_2) \gamma_\mu u(P_1,S_1)$ and 
$
\bar u(P_2,S_2) \sigma_{\mu\nu} P^\nu_- u(P_1,S_1)/M$,
respectively.\\ 
$f_a({\cal Z}, {\cal P}_i {\cal P}_j, x^2; \mu^2)$
are associated (renormalized) {\em two-variable distribution 
amplitudes}.

Correspondingly, non-forward matrix elements of operators 
with definite twist $\tau$ read:
\begin{eqnarray}
\hspace{-.7cm}
&&\langle P_2,S_2|
{\cal O}^{(\tau)}_{\Gamma}(\kappa x,-\kappa x)|P_1,S_1\rangle
\nonumber\\
\hspace{-.7cm}
&&=
{\cal K}_{\Gamma}^a({\cal P})
\int {D}{\cal Z}\; \e^{{\ii}\kappa (x{\cal P}){\cal Z}}\,
f^{(\tau)}_a({\cal Z}, {\cal P}_i {\cal P}_j, x^2; \mu^2 )
\nonumber\\
\hspace{-.7cm}
&&= {\cal P}^{(\tau)\Gamma'}_{~~\Gamma}\!\!(x,\pd_x) \,
{\cal K}_{\Gamma'}^a({\cal P})\!
\int\! {D}{\cal Z}\, \e^{\ii\kappa (x{\cal P}){\cal Z}}\,
f^{(\tau)}_a({\cal Z}; \mu^2 ),
\nonumber
\end{eqnarray}
where, in the first equality, using the twist decomposition of 
the operators the corresponding twist-$\tau$ DA's 
have been introduced and, in the second equality, using the Fourier 
representation, 
$
{\cal O}^{(\tau)}_{\Gamma}(\kappa x, -\kappa x)
= {\cal P}^{(\tau)\Gamma'}_{~~\Gamma}\!(x,\pd_x)
\int d^4 q 
\, \e^{{\ii}\kappa xq}
{\cal O}_{\Gamma'}(q), 
$ 
 and observing that the twist 
projections act on the exponentials only the {\em double distributions} 
$f^{(\tau)}_a({\cal Z};\mu^2)$ of definite twist $\tau$ not suffering
from any power corrections are introduced.

Therefore, the power corrections of the DA's  
$f_a^{(\tau)}({\cal Z}, {\cal P}_i {\cal P}_j, x^2; \mu^2 )$ 
are related to the DD's $f^{(\tau)}_{a}({\cal Z}; \mu^2 )$
according to: 
\begin{eqnarray}
\hspace{-.7cm}
&&f^{(\tau)}_a({\cal Z}, {\cal P}_i {\cal P}_j, x^2; \mu^2 )
= \big({\cal K}^{-1}({\cal P})\big)^{\Gamma}_a
\nonumber\\
\hspace{-.7cm}
&&\times
\bigg(\!\e^{-{\ii}\kappa (x{\cal P}){\cal Z}}\,
{\cal P}^{(\tau)\Gamma'}_{~\Gamma}
\! \e^{{\ii}\kappa (x{\cal P}){\cal Z}}\!\bigg)\,
{\cal K}_{\Gamma'}^{a'}({\cal P})\,
f^{(\tau)}_{a'}({\cal Z}; \mu^2 )
\nonumber\\
\hspace{-.7cm}
&&\equiv
{\cal F}^{(\tau)a'}_{a}\big((x{\cal PZ}), x^2 ({\cal PZ})^2\big)\,
f^{(\tau)}_{a'}({\cal Z}; \mu^2 ),
\nonumber
\end{eqnarray}
where ${\cal F}^{(\tau)a'}_{a}$ is uniquely related to the twist 
decomposition of the operators and contains any information about 
the power corrections of the twist-$\tau$ DD's.
Since they are independent of $x^2$ their renormalization properties
are already determined by the {\em light-cone operators} of definite 
twist. For the explicit structure of these operators, see, 
Ref.~\cite{GLR99,GL01}, and for the
non-forward matrix elements, see, Ref.~\cite{GLR01}.

The (quark) distributions $\hat f^{(\tau)}_{a}(z; \mu^2 )$
of definite twist $\tau$ are
obtained by taking forward matrix elements ($P_1=P_2=P, S_1=S_2=S$)
and restricting to the light-cone; they are obtained from the
DD's according to \cite{GLR01}
\begin{eqnarray}
\hat f^{(\tau)}_{a}(z; \mu^2 )
=\int d z_- \,f^{(\tau)}_{a}(z_+=z,z_-; \mu^2).
\nonumber
\end{eqnarray}

Now we present the above mentioned relations between the conventional
quark distributions of {\em dynamical} twist \cite{JJ92}, 
$e$, $f_1$, $f_4$, $g_1$, $g_T=g_2+g_1$, $g_3$, $h_1$, $h_L$ and $h_3$, 
and those of {\em geometric} twist, being denoted by
$E^{(3)}$, $F^{(2)}$, $F^{(4)}$, $G^{(2)}$, $G^{(3)}$, $G^{(4)}$, 
$H^{(2)}$, $H^{(3)}$ and $H^{(4)}$,
respectively (cf.~Ref.~\cite{GL01}): 
\begin{eqnarray}  
\hspace{-.7cm}
&&g_T(z)=G^{(3)}(z) + \int_z^1 \frac{d y}{y}
\Big(G^{(2)}-G^{(3)}\Big)\!\left(y\right),
\nonumber\\
\hspace{-.7cm}
&&g_3(z)=\int_z^1 \frac{d y}{y}\Big\{
\frac{1}{2}\Big(G^{(2)}-4G^{(3)}+3G^{(4)}\Big)\!\left(y\right)
\nonumber\\
\hspace{-.7cm}
&&\quad
+  \ln \Big(\frac{z}{y}\Big)
\Big(G^{(2)}-2G^{(3)}+G^{(4)}\Big)\!
\Big\} -\frac{1}{2}G^{(4)}(z) ,
\nonumber\\
\hspace{-.7cm}
&&f_4(z)=\frac{1}{2}F^{(4)}(z) + \frac{1}{2}\int_z^1 \frac{d y}{y}
\Big(F^{(2)}-F^{(4)}\Big)\!\left(y\right),
\nonumber\\
\hspace{-.7cm}
&&h_L(z)=H^{(3)}(z) + 2z \int_z^1 \frac{d y}{y^2}
\Big(H^{(2)}-H^{(3)}\Big)\!\left(y\right),
\nonumber\\
\hspace{-.7cm}
&&h_3(z)=\frac{1}{2}H^{(4)}(z)+ \int_z^1 \frac{d y}{y}\Big\{
\Big(H^{(2)}-H^{(4)}\Big)\!\left(y\right)
\nonumber\\
\hspace{-.7cm}
&&\quad
-\frac{z}{y}\Big(H^{(2)}-H^{(3)}\Big) 
-\frac{1}{2}\delta\Big(\frac{z}{y}\Big)\Big(H^{(3)}-H^{(4)}\Big)\!
\Big\}.
\nonumber
\end{eqnarray}
For leading twist both types of PD's coincide:
$g_1(z)=G^{(2)}(z),$
$f_1(z)=F^{(2)}(z),$
$e(z)=E^{(3)}(z),$ and
$h_1(z)=H^{(2)}(z)$. The corresponding inverse relations are:
\begin{eqnarray}
\hspace{-.7cm}
&&G^{(3)}(z)=g_T(z) +\frac{1}{z} \int_z^1 d y
\big(g_T-g_1\big)\!\left(y\right),
\nonumber\\
\hspace{-.7cm}
&&
G^{(4)}(z)=
-\Big\{\frac{1}{z^2} \int_z^1 d y\, y \Big[
\big(6g_3+4g_T-g_1\big)\!\left(y\right)
\nonumber\\
\hspace{-.7cm}
&&\qquad
+ 
\Big(1-\frac{z}{y}\Big)
\big(2g_3+4g_T-3g_1\big)\!\left(y\right)\Big]
+2g_3(z) \Big\},
\nonumber\\
\hspace{-.7cm}
&&F^{(4)}(z)=2f_4(z) + \frac{1}{z}\int_z^1 d y
\big(2f_4-f_1\big)\!\left(y\right),
\nonumber\\
\hspace{-.7cm}
&&H^{(3)}(z)=h_L(z) + \frac{2}{z}\int_z^1 d y
\big(h_L-h_1\big)\!\left(y\right)\,
\nonumber\\
\hspace{-.7cm}
&&H^{(4)}(z)=2\Big\{h_3(z) 
+\frac{1}{z^2} \int_z^1 d y\, y \Big[
\big(2h_3-h_L\big)\!\left(y\right)
\nonumber\\
\hspace{-.7cm}
&&\qquad\qquad\qquad\qquad
- 
\Big(1-\frac{z}{y}\Big)
\big(h_L-h_1\big)\!\left(y\right)\Big]\Big\}.
\nonumber
\end{eqnarray}

The {\em direct relations} define the decomposition of the conventional
PD's into their parts of definit geometric
twist. They may lead to Wandzura-Wilczek--like relations.
For example, the twist-2 part of $g_2(z)$ obviously coincides
with the well-known WW--relation, 
$g^{(\tau=2)}_2(z)\equiv g_2^{WW}(z) 
=-g_1(z) + \int_z^1 \frac{d y}{y}\,g_1(y)$,
and for $g_3(z)$ one gets two WW--like relations,
\begin{eqnarray}
\hspace{-.7cm}
&&g^{(\tau=2)}_3(z)=\int_z^1\frac{d y}{y}\Big(\frac{1}{2}
+\ln\frac{z}{y}\Big)g_1(y),
\nonumber\\
\hspace{-.7cm}
&&g^{(\tau=3)}_3(z)=-2\int_z^1 \frac{d y}{y}\Big(1+\ln\frac{z}{y}\Big)
\nonumber\\
\hspace{-.7cm}
&& \qquad\qquad\qquad\qquad
\times
\Big(g_T(y)+\frac{1}{y}\int_y^1 d u\, g_2(u)\Big).
\nonumber	
\end{eqnarray}
Analogously for the other distributions \cite{GL01}. 
From their lowest moments one obtains Burkhardt-Cottingham--like sum rules, 
e.g., for $g_3(z)$ from the first moments $n=0,1$ one gets \cite{GL01}:
\begin{eqnarray}
\hspace{-.7cm}
&&
\int_0^1 d z\, g_3(z) = - \frac{1}{2} \int_0^1  d z\, g_1(z),
\nonumber\\
\hspace{-.7cm}
&&
\int_0^1 d z\ z\ g_3(z) =
-\frac{1}{2} \int d z\, z\, (g_1 + 2 g_2)(z),
\nonumber
\end{eqnarray}
which 
may be of phenomenological advantage.
The {\em inverse relations} show that the PD's of definite twist are
uniquely determined by the conventional ones.


\begin{thebibliography}{9}
\bibitem{JJ92}
R. Jaffe and X. Ji, Nucl. Phys. B 375 (1992) 527.
\bibitem{GLR99} 
B. Geyer, M. Lazar and D. Robaschik, Nucl. Phys. B 559 (1999) 339,
B. Geyer and M. Lazar, Nucl. Phys. B 581 (2000) 341.
\bibitem{GL01} 
B. Geyer and M. Lazar, Phys. Rev. D 63 (2001) 094003.
\bibitem{GLR01} 
B. Geyer, M. Lazar and D. Robaschik, hep-ph/0108061;
to appear: Nucl. Phys. B.
\bibitem{L01}
M. Lazar, Phys. Lett. B 497 (2001) 62; JHEP01 (2001) 029;
M. Lazar and P. Ball, Phys. Lett. B 515 (2001) 131.
\end{thebibliography}
\end{document}